\documentclass[10.5pt,compsoc]{BD}
\usepackage{graphicx}
\usepackage{footmisc}
\usepackage{subfigure}
\usepackage{url}
\usepackage{multirow}
\usepackage[noadjust]{cite}
\usepackage{amsmath,amsthm}
\usepackage{amssymb,amsfonts}
\usepackage{hyperref}
\hypersetup{hidelinks=true}
\usepackage{booktabs}
\usepackage{color}
\usepackage{ccaption}
\usepackage{booktabs}
\usepackage{float}
\usepackage{fancyhdr}
\usepackage{caption}
\usepackage{xcolor,stfloats}
\usepackage{comment}
\graphicspath{{figures/}}
\usepackage{cuted}  
\usepackage{epstopdf}
\usepackage{bbding}
\definecolor{myred}{RGB}{236,114,112}
\headevenname{\zihao{-5}{\textbf{\emph{Big Data Mining and Analytics, January}}} 2018, 1(1): 000-000}%
\headoddname{{\sf Zeng et al.:}\quad {\textbf{\emph{A Zero-shot Explainable Doctor Ranking Framework with Large Language Models}}}}%

\newtheoremstyle{mystyle}{0pt}{0pt}{\normalfont}{1em}{\bf}{}{1em}{}
\theoremstyle{mystyle}
\renewcommand\figurename{Fig.~}

\newcommand{\upcite}[1]{\textsuperscript{\cite{#1}}}

\newcommand{\nop}[1]{}

\makeatletter
\renewcommand{\@biblabel}[1]{[#1]\hfill}
\makeatother

\makeatletter

\renewenvironment{thebibliography}[1]{%
    \begin{oldthebibliography}{#1}%
    \zihao{5-}%
    \setlength{\parskip}{0pt}%
    \setlength{\itemsep}{-0.1em}
    \setlength{\parsep}{0pt}%
}{%
    \end{oldthebibliography}%
}
\makeatother

\begin{document}

\thispagestyle{empty}

\hyphenpenalty=50000

\makeatletter
\newcommand\mysmall{\@setfontsize\mysmall{7}{9.5}}

\newenvironment{tablehere}
  {\def\@captype{table}}
  {}
\newenvironment{figurehere}
  {\def\@captype{figure}}
  {}

\thispagestyle{plain}%
\thispagestyle{empty}%

\let\temp\footnote
\renewcommand \footnote[1]{\temp{\zihao{-5}#1}}
{}
\vspace*{-40pt}
\noindent{\zihao{5-}\textbf{\scalebox{0.95}[1.0]{\makebox[5.9cm][s]
{BIG\hfill  DATA \hfill MINING \hfill AND \hfill ANALYTICS}}}}


\vskip .2mm
{\normalsize
\textbf{
\hspace{-5mm}
\scalebox{1}[1.0]{\makebox[5.6cm][s]{%
DOI:~\hfill1\hfill0\hfill.\hfill2\hfill6\hfill5\hfill9\hfill9\hfill/\hfill B\hfill D\hfill M\hfill A\hfill .\hfill2\hfill0\hfill 2\hfill 5\hfill.\hfill9\hfill0\hfill2\hfill0\hfill0\hfill 9\hfill 8}}}}

\vskip .2mm\noindent
{\zihao{5-}\textbf{\scalebox{1}[1.0]{\makebox[5.6cm][s]{%
V\hspace{0.4pt}o\hspace{0.4pt}l\hspace{0.4pt}u\hspace{0.4pt}m\hspace{0.4pt}%
e\hspace{0.4em}1\hspace{0.4pt},\hspace{0.8em}N\hspace{0.4pt}u\hspace{0.4pt}%
m\hspace{0.4pt}b\hspace{0.4pt}e\hspace{0.4pt}r\hspace{0.4em}1,\hspace{0.8em}%
J\hspace{0.4pt}a\hspace{0.4pt}n\hspace{0.4pt}u\hspace{0.4pt}a\hspace{0.4pt}%
\hspace{0.4pt}r\hspace{0.4pt}y\hspace{0.4em}2\hspace{0.4pt}0\hspace{0.4pt}1\hspace{0.4pt}8}}}}

\vskip .2mm\noindent
{\zihao{5-}\textbf{\scalebox{1}[1.0]{\makebox[5.6cm][s]{%
\color{white}{V\hfill o\hfill l\hfill u\hfill m\hfill%
e\hspace{0.356em}1,\hspace{0.356em}N\hfill u\hfill%
m\hfill b\hfill e\hfill r\hspace{0.356em}1,\hspace{0.356em}%
S\hfill e\hfill p\hfill t\hfill e\hfill%
m\hfill b\hfill e\hfil lr\hspace{0.356em}2\hfill0\hfill1\hfill8}}}}}\\

\begin{strip}
{\center
{\zihao{3}\textbf{
A Zero-shot Explainable Doctor Ranking Framework with Large Language Models}}
\vskip 9mm}

{\center {\sf \zihao{5}
Ziyang Zeng, Dongyuan Li, and  Yuqing Yang$^*$
}
\vskip 5mm}

\centering{
\begin{tabular}{p{160mm}}

{\zihao{-5}
\linespread{1.6667} %
\noindent
\bf{Abstract:} {\sf
Online medical service provides patients convenient access to doctors, but effectively ranking doctors based on specific medical needs remains challenging.
Current ranking approaches typically lack the interpretability crucial for patient trust and informed decision-making.
Additionally, the scarcity of standardized benchmarks and labeled data for supervised learning impedes progress in expertise-aware doctor ranking.
To address these challenges, we propose an explainable ranking framework for doctor ranking powered by large language models in a zero-shot setting. 
Our framework dynamically generates disease-specific ranking criteria to guide the large language model in assessing doctor relevance with transparency and consistency. 
It further enhances interpretability by generating step-by-step rationales for its ranking decisions, improving the overall explainability of the information retrieval process. 
To support rigorous evaluation, we built and released DrRank, a novel expertise-driven dataset comprising 38 disease-treatment pairs and 4,325 doctor profiles.
On this benchmark, our framework significantly outperforms the strongest baseline by +6.45 NDCG@10. Comprehensive analyses also show our framework is fair across disease types, patient gender, and geographic regions. 
Furthermore, verification by medical experts confirms the reliability and interpretability of our approach, reinforcing its potential for trustworthy, real-world doctor recommendation.
To demonstrate its broader applicability, we validate our framework on two datasets from BEIR benchmark, where it again achieves superior performance.
The code and associated data are available at: {\url{https://github.com/YangLab-BUPT/DrRank}}.
}
}

\end{tabular}
}
\vskip 6mm

\vskip -3mm
\zihao{6}\end{strip}

\thispagestyle{plain}%
\thispagestyle{empty}%
\makeatother
\pagestyle{tstheadings}

\begin{figure}[b]
\vskip -6mm
\begin{tabular}{p{44mm}}
\toprule\\
\end{tabular}
\vskip -4.5mm
\noindent
\setlength{\tabcolsep}{1pt}
\begin{tabular}{p{1.5mm}p{79.5mm}}
\\
$\bullet$& Ziyang Zeng and Yuqing Yang are with School of Information and  Communication Engineering, Beijing University of Posts and Telecommunications, Beijing 100876, China. Email: ziyang1060@bupt.edu.cn; yangyuqing@bupt.edu.cn \\
$\bullet$& Dongyuan Li is with School of Artificial Intelligence, Beijing University of Posts and Telecommunications, Beijing 100876, China. Email: li\_dongyuan@bupt.edu.cn \\
$\sf{*}$&
Yuqing Yang is the corresponding author. \\
          &          Manuscript received: 2025-Apr-19;
          accepted: 2025-Sep-08

\end{tabular}
\end{figure}\zihao{5}

\section{Introduction}
\label{s:introduction}
\noindent
The rapid development of internet medicine has greatly enhanced patients' access to trans-regional diagnostic and treatment services~\upcite{internetmedicine}.
Through professional online healthcare platforms, patients can search for and choose experienced doctors based on their suspected conditions and treatment needs, such as medication, surgery, or radiotherapy.
However, for patients with limited medical knowledge, selecting an appropriate doctor from a vast pool of professionals remains a significant challenge.

To improve the quality of online healthcare service, a key task is doctor ranking—that is, ranking doctors from a large candidate pool based on their relevance to a patient’s medical needs~\upcite{jiang2014finddoctor,guo2016doctor}.
This task has attracted growing attention in recent studies.
Early approaches typically rely on simple statistical indicators, such as the number of consultations or patient reviews~\upcite{Hu2018NewDR,Rahman2018DRanking}.
However, these metrics do not adequately reflect a doctor's professional expertise, often resulting in suboptimal rankings.
More recent methods adopt machine learning techniques to improve ranking accuracy~\upcite{Singh2023DRsimple,Chen2023DoctorsRT}.
Nevertheless, these models generally depend on labeled data for supervised learning, which is challenging to obtain in the medical domain due to the high cost of expert annotation.
Some studies attempt to construct supervised signals using proxy objectives—such as whether a consultation was completed~\upcite{KDD2022DRDialogue} or whether the doctor responded to an inquiry~\upcite{ACL2022DRDialogue}.
However, such heuristic signals often introduce commercial bias, shifting focus away from professional competence and compromising the reliability of ranking results for patients seeking domain-specific expertise.
Despite these advancements, existing methods largely overlook the explainability of doctor ranking, which is critical for building trustworthy recommendation systems~\upcite{rec_explanation}.
Meanwhile, {some commercial medical platforms—for example Haodf Online\footnote{\url{https://www.haodf.com}}—have released department-specific doctor rankings, but without disclosing the internal mechanisms behind the rankings.
In high-stakes domains such as healthcare, providing transparent explanations is indispensable to foster patient trust and secure acceptance of doctor rankings.

\begin{figure*}[!t]
\centerline{\includegraphics[width=\textwidth]{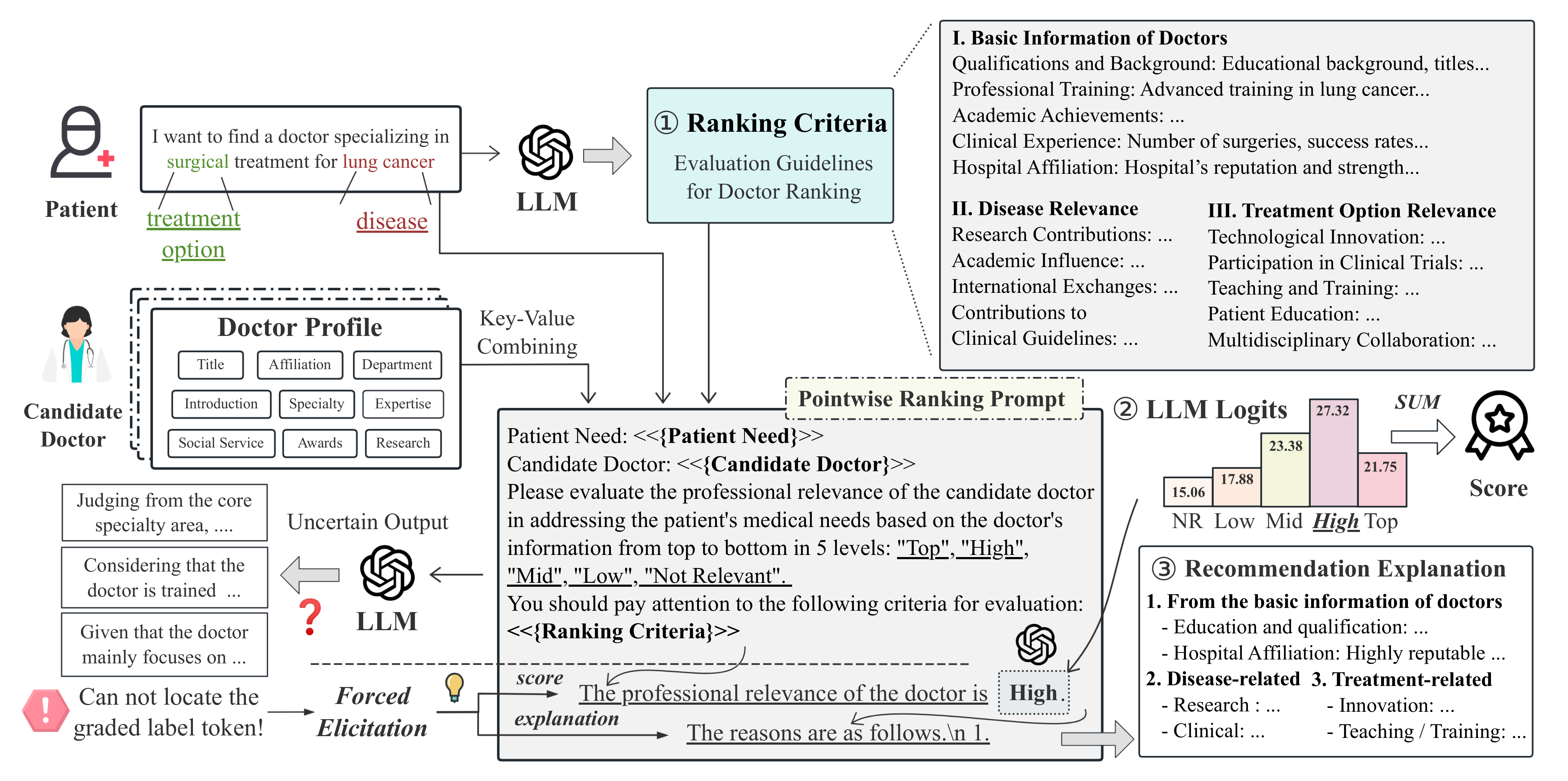}}
\caption{
An overview of our proposed zero-shot, large language model-based, explainable doctor ranking framework.
}
\label{fig:framework}
\end{figure*}

Large language models (LLMs), backed by extensive pre-training on large-scale textual data, have demonstrated impressive capabilities in performing text ranking tasks based on user-defined prompts~\upcite{pointwise2score,listwise2rankgpt,pairwise1,setwise1}.
Their remarkable generalization abilities and embedded medical knowledge make them highly suitable for the doctor ranking task, especially in data-scarce environments.
Inspired by this, we propose a novel large language model (LLM)-based doctor ranking framework, as illustrated in Fig.~\ref{fig:framework}, which ranks doctors according to their relevance to specific disease-treatment pairs.
The framework operates in a zero-shot setting, thereby eliminating the need for large-scale labeled training data.
To ensure efficiency, we adopt a pointwise ranking strategy, where the LLM predicts fine-grained relevance labels for individual doctor profiles.
These precise relevance labels help the model distinguish nuanced differences in professional expertise across a variety of disease-treatment scenarios.
To further enhance the consistency and interpretability of doctor ranking, we prompt the LLM to generate disease-specific ranking criteria, which serve as structured guidelines for relevance evaluation during the ranking decision-making process.
After generating relevance labels, the LLM is further prompted to produce step-by-step rationales, which serve as explanations for its ranking decisions.
We observe that the stochastic behavior of LLMs may affect both the relevance estimation and rationale generation.
To address this, we design a forced elicitation mechanism that encourages the LLM to consistently produce both relevance scores and their corresponding justifications in a controlled manner.

To enable rigorous and objective evaluation of doctor ranking, we introduce DrRank, the first expertise-driven doctor ranking benchmark dataset covering 38 disease-treatment pairs and 4,325 doctor profiles.
Each doctor profile is constructed using rich multidimensional information, including personal introductions, areas of expertise, community engagement, honors and awards, and academic achievements.
Moreover, we collaborate with certified medical experts to annotate doctor relevance in relation to specific disease-treatment pairs, providing high-quality supervision for evaluation.

Experimental results demonstrate that LLMs with more than 7 billion parameters significantly outperform baseline models, especially when guided by the generated ranking criteria.  
We further analyze the distribution characteristics of relevance scores with respect to the ground-truth label, revealing how our framework operates in practice. In addition, we evaluate the fairness of our framework, with results showing a significant advantage in fairness across disease types and patient gender over baseline methods. 
Furthermore, through manual evaluation of the generated criteria and rationales, we highlight the framework's practical utility and trustworthiness, reinforcing its potential for real-world doctor recommendation.
This strong performance is not limited to the medical domain; our framework's robustness and generalization are also validated on two datasets from BEIR benchmark, where it again achieves impressive zero-shot performance, confirming its broader applicability as a general-purpose ranking framework.

The main contributions of our work can be summarized as follows:
\begin{enumerate}
  \item We propose a zero-shot, LLM-based explainable doctor ranking framework that effectively addresses the challenge of limited labeled data in medical domains.
  \item We construct and release DrRank, the first expertise-driven benchmark dataset for objective evaluation of doctor ranking systems.
  \item Through extensive experiments, we demonstrated the accuracy, fairness, interpretability and generalizability of our framework, and we have released our code to ensure reproducibility.
\end{enumerate}

\section{Related Work}
\label{s:Related Work}
\subsection{Doctor Recommendation}
\label{rw:DR}
\noindent
Doctor recommendation aims to assist patients in identifying the most suitable medical professionals.
Unlike conventional recommendation systems that model user preferences based on historical behavior~\upcite{rec_survey}, doctor recommendation and ranking often rely on limited patient information—typically derived from brief chief complaints—due to privacy concerns.
Doctor modeling is another pivotal research focus in this domain.
Early approaches construct doctor profiles based on readily available attributes from online healthcare platforms~\upcite{Singh2023DRsimple,Chen2023DoctorsRT}, such as years of experience, consultation fees, and user reviews.
More recent studies leverage dialogue data between doctors and patients to enhance doctor representations and investigate how past interactions influence model predictions~\upcite{KDD2022DRDialogue,ACL2022DRDialogue}.
Other works explore personalized recommendation strategies to better align patients with doctors based on individual preferences and healthcare needs~\upcite{Ju2021DoctorRM,Shambour2024DR}.
In addition, Che et al.~\upcite{Che2021ADR} address the integration of heterogeneous data across multiple healthcare platforms, proposing methods to unify and utilize diverse sources of information.

Doctor ranking is typically considered a component of broader doctor recommendation systems. 
In this work, we focus exclusively on this aspect, aiming to generate accurate and explainable rankings that can serve as a foundation for building trustworthy recommendation systems in real-world healthcare platforms.
Yet, the absence of a standardized benchmark in this field makes it difficult to conduct fair comparisons across different approaches.
To address this gap, we present a professionally annotated doctor ranking dataset for the research community and benchmark general-purpose ranking models~\upcite{textranking}—including cross-encoder and bi-encoder architectures—as zero-shot baselines for future research.

\subsection{Large Language Models in Zero-shot Text Ranking}
\label{sec:llm_ranking}
\noindent
Shifting from tuning-based learning to rank driven by pretrained models~\upcite{textranking} such as MonoT5~\upcite{monoT5} and RankT5~\upcite{zhuang2023rankt5}, there is an emerging line of research exploring how to utilize general-purpose LLMs for zero-shot text ranking in out-of-domain scenarios.
The current LLM-based ranking methods can be mainly divided into the following three categories: 
(1) \textit{Pointwise} strategy utilizes LLMs to measure the relevance between a query and a single passage by query generation~\upcite{pointwise1query} or relevance generation~\upcite{pointwise1score,pointwise2score}.
(2) \textit{Pairwise} strategy prompts LLMs to determine which of two passages is more relevant to a query and aggregates a final relevance score for each passage from all pair comparisons, where efficient sorting algorithms, such as heapsort, can be used to speed up the ranking process particularly in top-k ranking~\upcite{pairwise1}. Setwise strategy enhances the efficiency of pairwise comparisons by selecting the most relevant passage from multiple candidates~\upcite{setwise1}.
(3) \textit{Listwise} strategy instructs LLMs to generate a permutation of passages, arranged in descending order of relevance to a query~\upcite{listwise2rankgpt}. Due to the limited input length of LLMs, a sliding window approach is used to rerank a subset of candidate passages each time, moving the window from back to front to cover all candidates. 
In this work, we leverage the zero-shot pointwise ranking capabilities of LLMs to address the challenge of training doctor ranking models with limited labeled data.

\subsection{Explainable Ranking}
\noindent
In recent years, the field of explainable ranking has garnered significant attention, aiming to enhance the transparency and trustworthiness of ranking models.
Existing approaches can be broadly categorized into two lines of work: intrinsically interpretable ranking models and natural language-based explanations.
The first line focuses on designing models that are inherently interpretable, such as additive models~\upcite{Additive}, which provide clear insights into how individual human-engineered ranking features~\upcite{Blackbox} contribute to ranking decisions. 
However, as ranking tasks grow increasingly complex, the performance of such interpretable models is often limited by their simplistic architectures.
A more promising direction involves leveraging natural language generation to produce human-understandable explanations for ranking outcomes.
For example, ExaRanker~\upcite{ExaRanker} incorporates synthetic explanations generated by LLMs to enhance both training efficiency and the interpretability of sequence-to-sequence ranking models.
LiEGe~\upcite{ListwiseExplaination} jointly explains all documents in a ranked list by learning semantic representations at two levels of granularity—documents and their constituent tokens.
A work most related to ours is MCRanker~\upcite{adaptivemcranker}, which enhances LLM-based pointwise ranking through multi-perspective criteria.
This approach can be viewed as a more complex variant of ours, involving multiple rounds of LLM inference and result ensembling from different perspectives.
While it potentially offers improved performance, it significantly increases computational overhead and system complexity.

However, prior works typically focus on either the ranking process or the ranking output, and rarely integrate both into a unified explainable framework. 
In contrast, our approach jointly models ranking and explanation generation, using LLMs to produce both explicit criteria and step-by-step rationales. 
This unified framework simultaneously enhances the interpretability of both the decision-making process and the final outcomes, making the overall pipeline more transparent and trustworthy.

\section{DrRank Dataset}
\label{sec:dataset}
\noindent
In this section, we elaborate on the construction of the DrRank dataset and provide an overview of its statistical information.

\begin{figure}[!t]
\centerline{\includegraphics[width=\columnwidth]{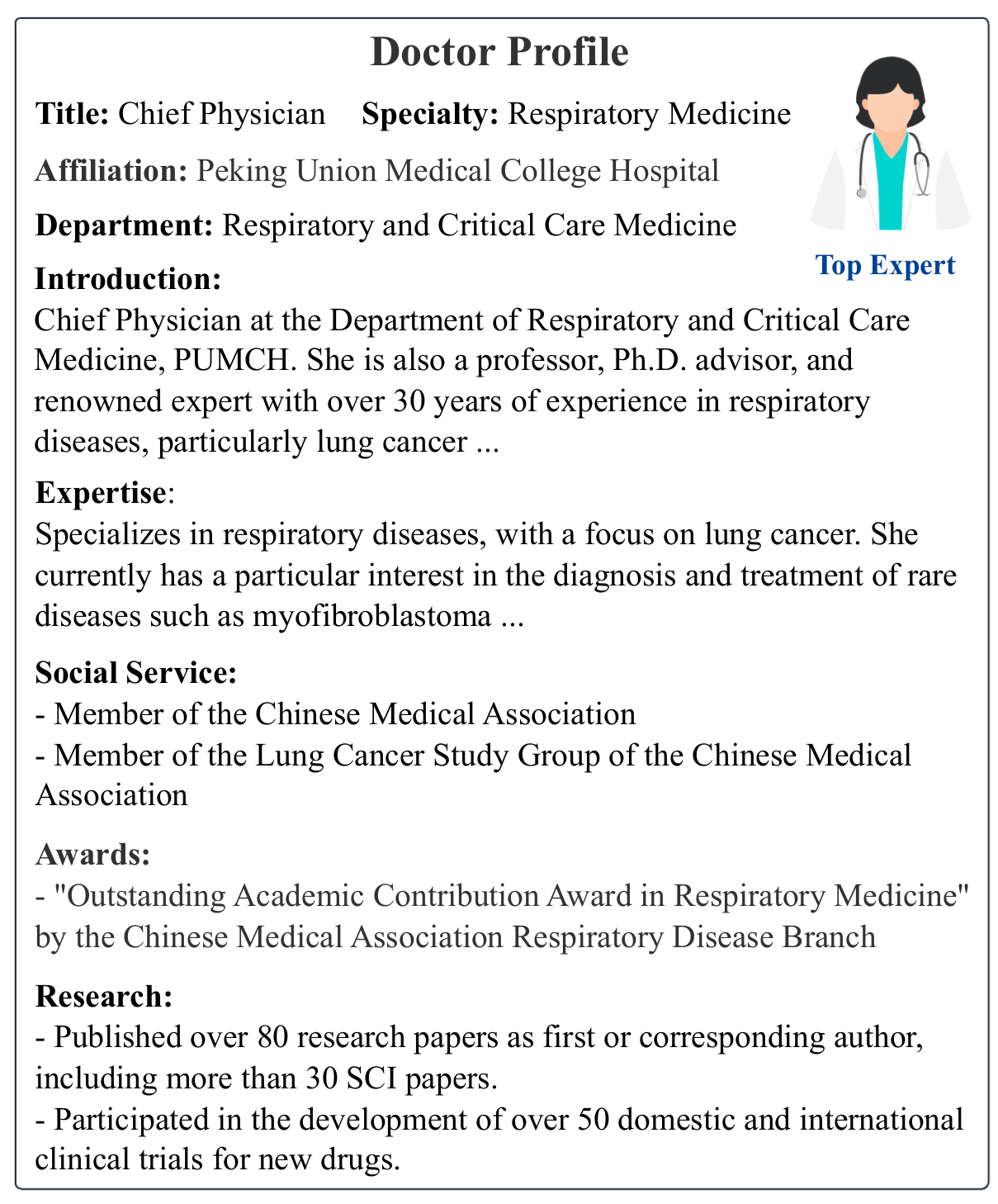}}
\caption{An example of doctor profiles used in the DrRank dataset.}
\label{fig:doctor}
\end{figure}

\subsection{Data Collection and Preprocessing}
\noindent
The doctor profiles used in this study is primarily sourced from Haodf Online, one of the biggest online healthcare platforms in China.
Each doctor profile comprises 9 fields: ``Title", ``Specialty", ``Affiliation", ``Department", ``Introduction", ``Expertise", ``Social Service", ``Awards", and ``Research", as depicted in Fig.~\ref{fig:doctor}. 
Over 600,000 doctor profiles were collected in total, forming a comprehensive resource for analysis.
To proceed, we first selected a range of common cancers and critical illnesses, along with their typical treatment options, resulting a set of 38 disease-treatment pairs. 
For each pair, such as lung cancer and surgical treatment, we employed the open-sourced gte-multilingual-reranker-base\footnote{\url{https://huggingface.co/Alibaba-NLP/gte-multilingual-reranker-base}} model~\upcite{zhang2024mgte} to assess the relevance of the introductions from the doctor profiles in our extensive pool to the disease.
This process yielded an initial list of 100 candidate doctors for each disease-treatment pair.

\subsection{Data Annotation}
\label{sec:da}
\noindent
The annotation team was composed of clinical experts from Peking Union Medical College Hospital—consistently ranked as China’s top hospital in comprehensive evaluations. 
For each disease–treatment pair, two specialists in the relevant field were interviewed for approximately one hour. 
During these interviews, they reviewed our preliminary list of doctors, assessed each individual’s suitability, and recommended any additional professionals who, based on real-world practice, should also be included.
Evaluation criteria encompassed disease relevance, hospital tier, physician title, positions held within professional societies, research accomplishments, and honors received. 
Each doctor was then assigned a relevance score from 0 to 5 (5 = highest relevance; 0 = irrelevant). 
In practice, these scores correspond to six tiers: leading experts, nationally recognized specialists, provincially acknowledged experts, municipal-level authorities, general specialists, and irrelevant physicians.
Doctors whose scores agreed between both reviewers were retained outright; whenever the two experts’ scores diverged, a third independent specialist adjudicated to determine the final score.
It is worth noting that the annotated dataset encompasses professionals from a wide range of institutions rather than focusing solely on top-tier hospitals.

\begin{figure}[!t]
\centerline{\includegraphics[width=\columnwidth]{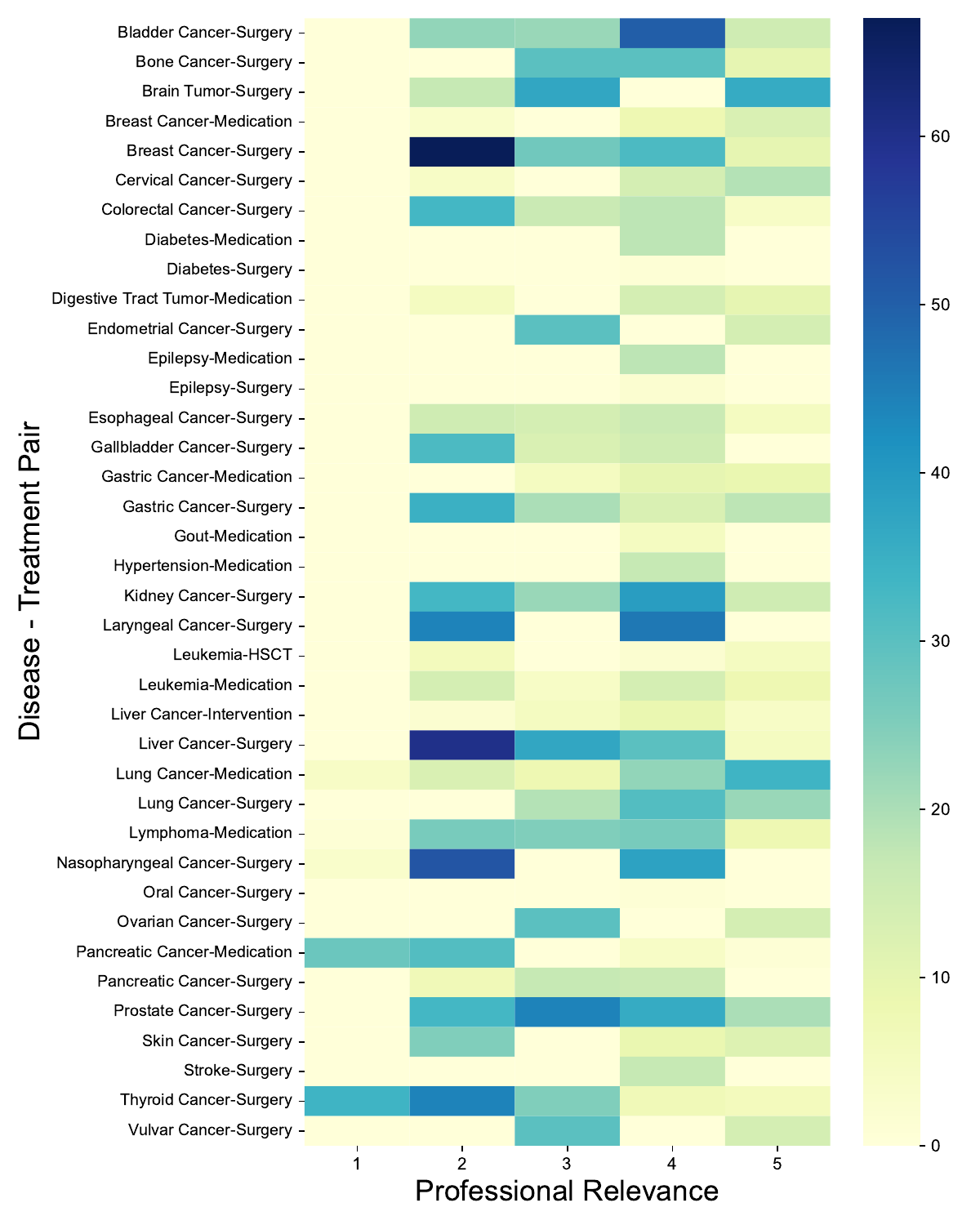}}
\caption{
Heatmap of the annotated professional relevance distribution across positive samples for different disease-treatment pairs in the DrRank dataset.
}
\label{fig:score_heatmap}
\end{figure}

\subsection{Hard Negative Mining}
\label{sec:hnm}
\noindent
To balance the number of positive and negative samples for each disease-treatment pair and enhance the complexity of dataset, we conducted hard negative mining.
We considered doctors with a relevance score greater than 1 as positive samples and those with a relevance score of 0 as negative samples. 
Statistical analysis of the doctor profiles in the annotated positive samples revealed that they tend to be relatively long. Therefore, we filtered the profiles to include only those with a total length exceeding 1,024 \footnote{Unless otherwise specified, all lengths mentioned in this work are defined based on the tokenizer inherent to the Qwen2.5 series.}. 
This filtering process resulted in a subset of 5,000 unique doctors with sufficiently rich profile information.
For each disease-treatment pair, we utilized the gte-multilingual-reranker-base again to rank the filtered set of doctors based on the relevance score between the disease and the doctor's introduction.
To minimize false negatives, we excluded the top 1\% of doctors from the ranked list. 
Then, from the remaining list, we iteratively added negative samples not initially included, in descending order of reranker scores, until we achieved a 1:1 ratio of positive to negative samples for each disease-treatment pair.
To further introduce more challenging negative samples, we randomly replaced 30\% of the previously selected negative samples with positive samples from other pairs. 
Specifically, we regarded the positive samples with relevance scores of 4 and 5 from other disease-treatment pairs as negative samples for the current pair.
Finally, we conducted a manual review to ensure that the all selected negative samples for each disease-treatment pair were indeed unrelated to it.

\subsection{Dataset Statistics}
\noindent
The DrRank dataset, constructed through the aforementioned process, includes 30 disease types, comprising 23 cancers and 7 other critical illnesses.
Each disease is linked to one or more treatment options such as medication, surgery, radiotherapy, and, in the case of leukemia, Hematopoietic Stem Cell Transplantation (HSCT). 
Although most diseases are associated with a single treatment option, some involve multiple treatments, resulting in a total of 38 unique disease-treatment pairs. 
Our dataset includes 4325 doctor profiles, of which 2,593 are unique. 
Some doctors are listed in multiple disease-treatment pairs due to their expertise spanning a variety of conditions and treatments.
Fig.~\ref{fig:score_heatmap} illustrates the distribution of annotated professional relevance scores among positive samples across all the disease-treatment pairs in the DrRank dataset, where the x-axis represents the professional relevance (ranging from 1 to 5) and the y-axis lists various disease-treatment pairs.
Furthermore, in Fig.~\ref{fig:length_distribution}, the x-axis represents the length of tokenized doctor profiles, and the y-axis represents the cumulative percentage of sequences. The vertical dashed lines at 1024, 2048, and 4096 tokens indicate key truncation thresholds. 
Each of these thresholds is annotated with the percentage of the DrRank dataset that is covered up to that token count.
The distribution reveals substantial variability in the length of doctor profiles, with the average length in the DrRank dataset exceeding that of our initial doctor pool.
These findings underscore the complexity of the doctor ranking task in real-world scenarios, given the significant variation in the volume and diversity of available information. 

\begin{figure}[!t]
\centerline{\includegraphics[width=\columnwidth]{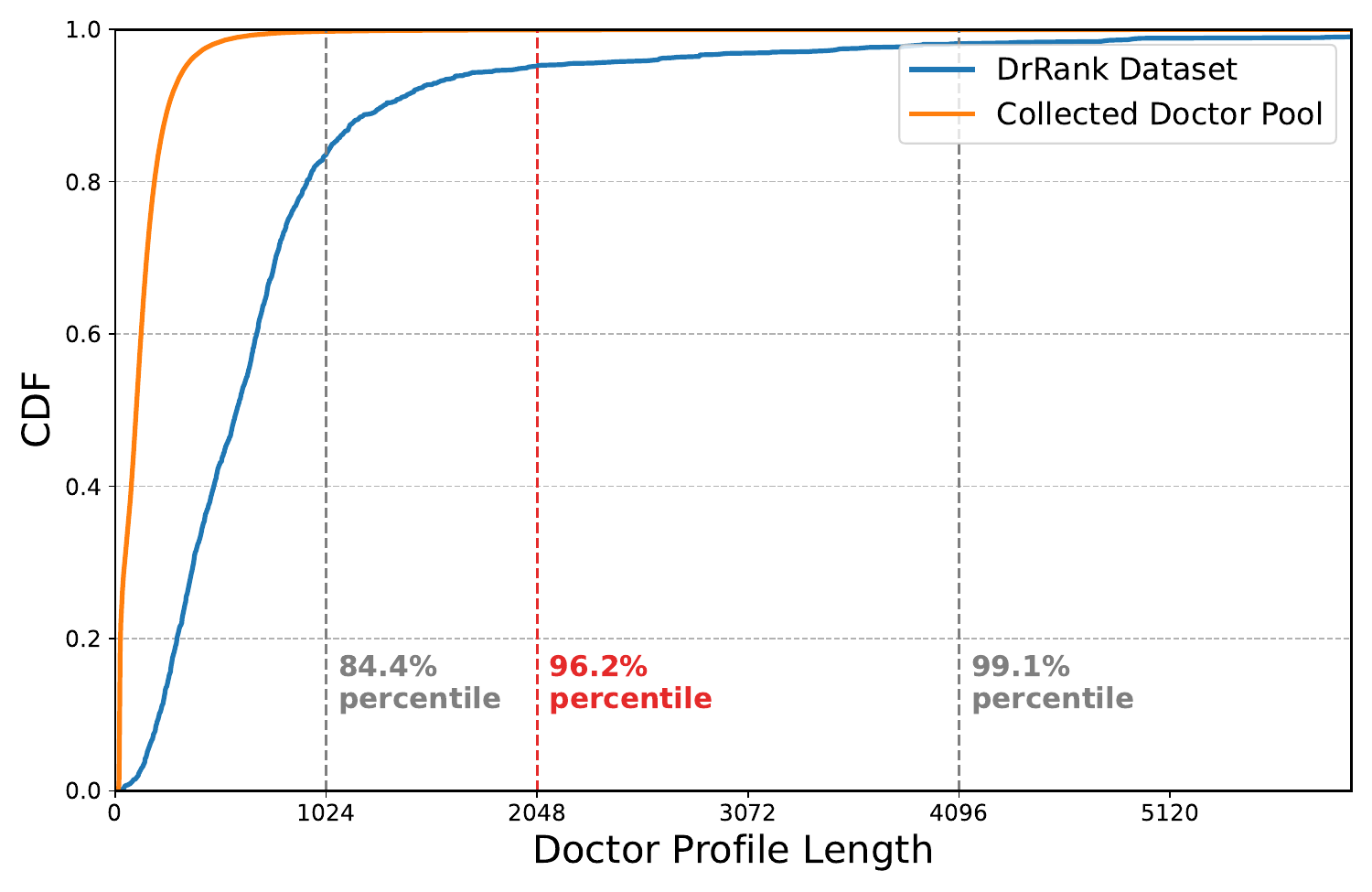}}
\caption{
Cumulative distribution functions (CDF) of doctor profile lengths from two datasets: DrRank and Collected Doctor Pool.
}
\label{fig:length_distribution}
\end{figure}

\section{Methodology}
\subsection{Problem Formulation}
We construct the query set \(\mathcal{Q}\) by pairing specific diseases with treatment options to simulate real patient medical needs, such as ``I want to find a doctor specializing in surgical treatment for lung cancer".
The set of candidate doctors is denoted as \(\mathcal{D}\). 
Each doctor \(d \in \mathcal{D}\) is represented by a detailed text which is constructed by sequentially combining key-value pairs from the doctor’s profile.
Each key-value pair consists of a key (e.g., ``Title") and its corresponding value (e.g., ``Chief Physician"). Similarly, another key could be ``Specialty" with the value ``Respiratory Medicine". We can combine the aforementioned two key-value pairs in the following text description: ``Title: Chief Physician \textbackslash n Specialty: Respiratory Medicine", where ``\textbackslash n" represents a newline character used to separate different key-value pairs.
Further details on the keys in doctor profiles are provided in Section \ref{sec:dataset}. Along  this line, the problem of doctor ranking can be formulated as follows: 

\newtheorem{definition}{Definition}
\begin{definition}[Doctor Ranking]
Given a patient's medical need \(q \in \mathcal{Q}\) and a candidate doctor \(d \in \mathcal{D}\), the goal of doctor ranking is to evaluate the professional relevance of doctors to patients' specific medical needs, i.e., \(f(q, d) \rightarrow \mathbb{R}\).
\end{definition}

\subsection{LLM-based Doctor Ranking}
\noindent
Fig.~\ref{fig:framework} presents a comprehensive overview of our proposed ranking framework, which integrates several key components to ensure accurate, efficient, and explainable doctor ranking. 
Below, we will elaborate on the details of our framework.

\subsubsection{Exploiting Efficient and Effective Pointwise Ranking}
\noindent
Although pairwise and listwise ranking methods have the capability to compare candidates relatively during inference time, their practical application in real-world scenarios remains extremely challenging. 
Pairwise comparisons suffer from quadratic complexity~\upcite{pairwise1}, incurring significant computational overhead.
Meanwhile, the sliding window mechanism employed by listwise method is inherently difficult to parallelize~\upcite{listwise2rankgpt}, resulting in substantial latency in large-scale data sorting tasks.
Additionally, both pairwise and listwise methods are susceptible to \textit{position bias}~\upcite{consistency_listwise,lost_middle,pairwise1}, where the order of candidates can influence the ranking outcomes. 
This bias introduces instability, severely compromising the credibility of ranking results, particularly in critical medical scenarios where precision and reliability are paramount.

In contrast, the pointwise ranking method excels with its high degree of parallelism, allowing for independent scoring of each candidate~\upcite{pointwise1query,pointwise1score}. 
A notable advantage of the pointwise method lies in the fact that it generates scores which are free from position bias. 
This ensures that the scores are stable, facilitating easy reproducibility of ranking results. 
Furthermore, by harnessing the robust language generation capabilities of LLMs, generating the scoring rationale for each candidate becomes straightforward, enhancing the transparency and trustworthiness of the ranking process.
In comparison, pairwise and listwise methods usually focus on relative comparisons between pairs or lists of candidates, making it difficult to extract and present individual, easily understandable rationales for each candidate.
Considering the aspects of parallel efficiency, score stability and interpretability, the pointwise ranking method emerges as the ideal choice for doctor ranking.

\subsubsection{Utilizing Fine-grained Labels to Optimize Distinction Relevance}
\noindent
Existing research utilizing LLMs as zero-shot pointwise rankers can be broadly categorized into two approaches: query generation~\upcite{pointwise1query} and relevance generation~\upcite{pointwise1score,pointwise2score}.
This work focuses on the relevance generation approach. 
In this approach, the LLM is prompted to determine whether a document is relevant to a query, using the logits of ``Yes" and ``No" tokens to derive a relevance score.
However, studies involving human subjects have demonstrated that binary options can sometimes lead to biased responses, whereas offering options with a higher level of granularity can produce more dependable outcomes~\upcite{binary_options}. 
Zhuang et al.~\upcite{pointwise2score} suggest incorporating detailed relevance labels into the prompt for pointwise LLM rankers, allowing them to better distinguish among documents with differing degrees of relevance to the query and thus generate a more precise ranking.

Inspired by these insights and aligned with the data‐annotation labels (Section \ref{sec:da}), we adopt a five‐level grading scheme—``Top", ``High", ``Mid", ``Low", and ``Not Relevant"—to evaluate each doctor’s professional relevance for addressing patients’ specific medical needs.
For each query-doctor pair $(q, {d_i})$, we instruct the LLM to evaluate their professional relevance by selecting from the provided graded labels, with the prompt $p$ shown in Fig.~\ref{fig:framework}.
We can obtain the corresponding $\mathrm{logit_k}$ for each graded label $l_k$ from the LLM output as \eqref{eq1}, which represents the raw confidence values of evaluation.

\begin{equation}\label{eq1}
\begin{split}
    \mathrm{logit_{i,k}}
    &= \mathrm{LLM}(l_k \mid p, q, d_i),\\
    \text{where } k &\in \{0,1,2,3,4\}.
\end{split}
\end{equation}
We can use these logits to derive the ranking score by calculating the expected value.
Firstly, we need to assign a score $s_k$ to each graded label $l_k$ . 
Through empirical investigation, we find that naively assigning $s_k = k$ (with 0 to 4 ordered from least to most relevance) already yields excellent performance.
Then, we can calculate the expected scoring value as follows:

\begin{equation}\label{eq2}
\begin{split}
    f(q, d_i) &= \sum_{k} \mathrm{prob_{i,k}}\cdot s_k,\\
    \text{where } \mathrm{prob_{i,k}} &= \frac{\exp(\mathrm{logit_{i,k}})}{\sum_{k'}\exp(\mathrm{logit_{i,k'}})}.
\end{split}
\end{equation}
In this context, the marginal probability $\mathrm{prob_{i,k}}$ can be interpreted as the likelihood of the doctor $d_i$'s relevance belonging to the graded label $l_k$.
In fact, this approach to finding the expected average is akin to sampling the outputs of LLMs multiple times~\upcite{pointwise2score}. 

In preliminary experiments, we observed an unexpected behavior: instruction-tuned LLMs, when prompted to generate a graded assessment, do not always produce outputs that directly correspond to our predefined 5-level label set.
Specifically, the first token or phrase generated by the model may deviate from the expected label space, introducing ambiguity and inconsistency when extracting the associated logits for subsequent ranking score computations.
This inconsistency undermines the reliability of pointwise scoring mechanisms that rely on matching model outputs to specific labels.
To address this issue, we propose a forced elicitation mechanism. 
This mechanism involves appending a fixed, semantically aligned labeling prefix to the end of each prompt to steer the model’s generation toward a constrained, predictable output space. 
For instance, by appending the phrase ``The professional relevance of the candidate doctor is '', we provide a strong inductive bias that encourages the model to follow with one of the target graded labels.
Empirically, we find that this prefix effectively anchors the model's generation, ensuring that the first generated token aligns exactly with our predefined label set. 
This alignment not only facilitates accurate logit extraction but also enhances the interpretability and consistency of model behavior across queries.
This forced elicitation strategy can be seen as a lightweight yet robust form of constrained decoding, implemented at the prompt design level without requiring modification of the LLM’s architecture or decoding algorithm.

\subsubsection{Generating Ranking Criteria as Standard Guidelines}
\noindent
LLMs are widely known as extremely intricate ``black-box'' systems~\upcite{llm_explain_survey}.
The iterative nature of next-token prediction results in ambiguous assessment criteria for zero-shot pointwise rankers~\upcite{adaptivemcranker}, leading to a lack of consistency and transparency in the scoring process.
Establishing clear criteria by which LLMs score doctors is crucial for building an accurate and explainable doctor ranking system. 
Recent studies on human annotation practices have emphasized the importance of a standardized annotation guideline for ensuring clear and consistent communication with annotators~\upcite{human_annotation1,human_annotation2}.

Inspired by these, we argue that standard ranking criteria should be provided for LLMs to ensure doctors are evaluated under a unified and transparent standard, thereby enhancing the consistency and accuracy of the final doctor ranking results.
To this end, we meticulously design a prompt (as shown in Fig.~\ref{fig:criteria}) to instruct LLMs to generate doctor ranking criteria for specific disease-treatment pairs.
Initially, we repeatedly prompt LLMs to generate multiple sets of ranking criteria for a specific disease-treatment pair. 
Subsequently, we manually inspect these criteria sets to select the best one, based on factors such as comprehensiveness, practicality, and overall format.
Using the optimal criteria as an example, we adopt a one-shot prompting approach to generate consistent ranking criteria for other disease-treatment pairs.
Our method dynamically integrates disease-specific ranking criteria into the ranking prompt, serving as contextual guidelines for the rankers' decisions.
These disease-specific factors are essential for aligning ranking results with patients' needs and ensuring that ranking decisions are based on the most relevant professional factors.

\begin{figure}[!t]
\centerline{\includegraphics[width=\columnwidth]{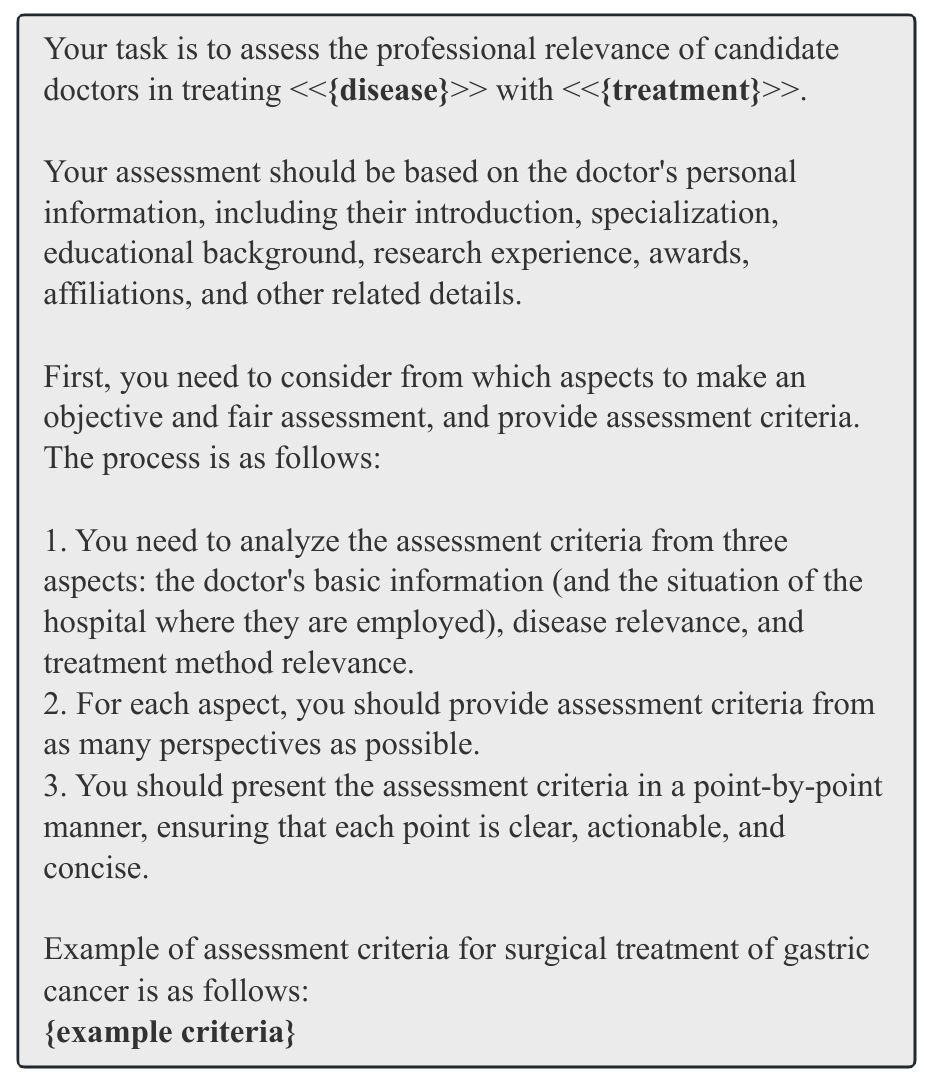}}
\caption{Prompt for ranking criteria generation.}
\label{fig:criteria}
\end{figure}

\subsubsection{Generating Rationales as Ranking Explanations}
\noindent
In the context of medical applications, particularly doctor ranking, providing transparent and comprehensible explanations is essential.
Patients often seek not only a ranked list of candidate doctors, but also an understanding of why a particular doctor is ranked higher than others. 
Such explanations help alleviate uncertainty and foster trust in the recommendation system~\upcite{rec_explanation}.

To address this need, we leverage the LLM itself (i.e., the ranker) to generate natural language rationales that serve as explanations for the doctor ranking outcomes.
After computing the logits corresponding to the predefined five-level graded labels, we select the label with the highest logit as the model's predicted assessment. 
This predicted label is then concatenated back into the original ranking prompt, forming a complete declarative evaluation of the doctor's professional relevance.
Next, we invoke the forced elicitation mechanism again—this time to elicit a rationale that justifies the selected label. 
Specifically, we append a predefined explanation prefix (i.e., ``.\textbackslash n\textbackslash n The reasons are as follows.\textbackslash n 1.'') to the prompt, which primes the model to generate a structured explanation in list format. 
This structure enhances readability and aligns with users' expectations for step-by-step reasoning.
By incorporating rationale generation into the ranking process, our framework produces not only interpretable rankings but also clear, human-understandable justifications. 
These explanations ultimately enhance user trust and satisfaction with the system’s doctor recommendation outcomes, supporting more informed healthcare decisions.

\section{Experiments}
\label{sec:exp}

\subsection{Experimental Settings}
\noindent
We evaluate the proposed framework on the DrRank dataset. 
We omit the dataset description in this section since it has been introduced in Section \ref{sec:dataset}. 
Other experimental settings will be described in the following parts.

\subsubsection{Baseline Methods}
\noindent
As discussed in Section~\ref{rw:DR}, related work in doctor recommendation faces significant evaluation hurdles due to the lack of standard benchmarks. 
This issue is further compounded by the unavailability of open-source data and models in many existing approaches~\upcite{Singh2023DRsimple,KDD2022DRDialogue,ACL2022DRDialogue}, alongside the intricate and diverse nature of real-world medical scenarios.
To advance research in doctor ranking, we contribute a professionally annotated dataset and benchmark a set of general-purpose ranking models, including bi-encoder and cross-encoder architectures, as zero-shot baselines.
In the bi-encoder setup, patient queries and doctor profile texts are independently encoded into dense vectors, and their cosine similarity is used as the ranking score. 
In contrast, the cross-encoder setup concatenates the patient query with structured doctor profile information and feeds the combined sequence into a reranker to compute a relevance score.
For the bi-encoder baselines, we include \texttt{bge-m3}~\upcite{bge_m3_1}, \texttt{jina-embeddings-v3}~\upcite{jinaembeddingv3}, \texttt{e5-mistral-7b-instruct}~\upcite{Wang2023ImprovingEmbeddingLLM}, and \texttt{gte-Qwen2-7B-instruct}~\upcite{gte_embedding}. 
For cross-encoders, we adopt \texttt{jina-reranker-v2-base}, \texttt{bce-reranker-base}, \texttt{bge-reranker-v2-m3}, and \texttt{bge-reranker-v2-gemma}.
These models have demonstrated strong performance on a variety of out-of-domain retrieval and ranking tasks. 
Since they are typically trained on general-domain datasets such as MS MARCO~\upcite{msmarco}, their performance on DrRank reflects their zero-shot generalization ability in the healthcare domain.
We also report results for pairwise and listwise LLM-based ranking (described in Section~\ref{sec:llm_ranking}), where each case is evaluated by jointly considering two or more candidate doctors. 
While pairwise and listwise approaches can deliver improved performance, they come at the expense of greater computational cost and markedly higher inference latency.
As such, these results are not directly comparable to pointwise methods. Instead, we present them as an upper bound to illustrate the potential of pairwise and listwise ranking, particularly when leveraging state-of-the-art LLMs such as \texttt{gpt-4o}.


\begin{table*}[h!]
\caption{
Model Performance on the DrRank dataset.
Model sizes of Bi-Encoder and Cross-Encoder are denoted as [parameters].
Model with specifically generated ranking criteria is marked with *.
Model with randomly assigned ranking criteria is marked with $^{+}$.
The top results in each group are highlighted in bold.
}
\centering
\begin{tabular}{cclcc}
\toprule
\multicolumn{2}{l}{Category} & \multicolumn{1}{l}{Model} & 
\multicolumn{1}{l}{NDCG@10} & \multicolumn{1}{l}{Recall@10}\\ 
\midrule
\multicolumn{2}{l}{\multirow{4}{*}{Bi-Encoder}} & bge-m3 [568M] & 63.49 & 29.00 \\
\multicolumn{2}{l}{} & jina-embeddings-v3 [572M] & 60.54 & 28.10 \\
\multicolumn{2}{l}{} & e5-mistral-7b-instruct [7.11B] & 69.21 & 29.98  \\
\multicolumn{2}{l}{} & gte-Qwen2-7B-instruct [7.61B] & \textbf{70.96} & \textbf{30.03} \\
\midrule
\multicolumn{2}{l}{\multirow{4}{*}{Cross-Encoder}} & jina-reranker-v2-base [278M] & 57.98 & 25.02 \\
\multicolumn{2}{l}{} & bce-reranker-base\_v1 [279M] & 62.86 & 28.37   \\
\multicolumn{2}{l}{} & bge-reranker-v2-m3 [568M] & 69.36 & 29.01   \\
\multicolumn{2}{l}{} & bge-reranker-v2-gemma [2.51B] & \textbf{69.59} & \textbf{29.91}   \\
\midrule
\multirow{19}{*}{LLM-based Ranker} & Pairwise & Qwen2.5-32B-Instruct & \textbf{78.13} & \textbf{34.17}   \\
\cmidrule(lr){2-5}
& \multirow{4}{*}{Listwise} & gemini-2.0-flash & 81.59 & 31.81   \\
&  & Qwen2.5-32B-Instruct & 82.39 & 31.86   \\
&  & claude-3-5-sonnet-20241022 & 83.20 & 32.07  \\
&  & gpt-4o-2024-08-06 & \textbf{84.24} & \textbf{32.21}   \\
\cmidrule(lr){2-5}
 & \multirow{14}{*}{Pointwise (Ours)} & Qwen2.5-0.5B-Instruct* & 45.01 & 22.16 \\
 &  & Qwen2.5-0.5B-Instruct & 55.18 & 24.47 \\
 &  & Qwen2.5-1.5B-Instruct* & 58.49 & 27.01 \\
 &  & Qwen2.5-1.5B-Instruct & 59.65 & 26.91 \\
 &  & Qwen2.5-3B-Instruct* & 65.22 & 27.95 \\
 &  & Qwen2.5-3B-Instruct & 64.28 & 27.74 \\
 &  & Qwen2.5-7B-Instruct* & 73.73 & 29.99 \\
 &  & Qwen2.5-7B-Instruct$^{+}$ & 72.16 & 29.37 \\
 &  & Qwen2.5-7B-Instruct & 71.44 & 29.27 \\
 &  & Qwen2.5-32B-Instruct* & \textbf{77.41} & \textbf{30.73} \\
 &  & Qwen2.5-32B-Instruct$^{+}$ & 76.61 & 30.40 \\
 &  & Qwen2.5-32B-Instruct & 76.57 & 30.32 \\
 &  & Baichuan-M1-14B-Instruct* & 74.67 & 29.82 \\
 &  & Baichuan-M1-14B-Instruct & 71.93 & 28.49  \\
\bottomrule
\end{tabular}
\label{table:main}
\end{table*}

\subsubsection{Implementation Details}
\noindent
Bi-encoder and cross-encoder models are used with their default configurations. 
For pairwise LLM-based ranking, we follow the implementation of PRP~\upcite{pairwise1} and use heapsort sorting algorithm for efficiency due to its guaranteed $O(N log N )$ computation complexity.
For listwise LLM-based ranking, we follow the implementation of RankGPT~\upcite{listwise2rankgpt}, adopting a sliding window strategy with a window size of 20 and a step size of 10.
To comprehensively evaluate the effectiveness of our proposed framework, we conducted experiments using the general-purpose instruction-tuned language models of the Qwen2.5 series~\upcite{qwen2025qwen25technicalreport}, ranging from 0.5B to 32B in size. These experiments allowed us to demonstrate the relationship between model scale and zero-shot ranking performance.
In addition to the Qwen2.5 series, we also evaluated \texttt{Baichuan-M1-14B-Instruct}~\upcite{baichuanm12025}, an open-source large language model developed specifically optimized for medical scenarios. 
While prior studies have shown that medical pretraining can enhance performance in clinical tasks such as question answering and dialogue generation~\upcite{medical_llm}, its application to medical ranking tasks remains underexplored.

Our framework was implemented using PyTorch 2.5.1 and Transformers 4.44.2 based on Python 3.10.15, and executed on a Linux machine equipped with 4 × NVIDIA A6000 GPUs and 256GB of memory.
To ensure the stability and consistency of LLM outputs, we adopt greedy decoding in all our experiments, resulting in deterministic model predictions.
We set a maximum length of 2,048 for the detailed text constructed from the doctor profile, truncating any content that exceeded this limit.
Furthermore, we configured a maximum length of 1,024 for the model-generated criteria and a maximum length of 512 for the evaluation rational.
Note that the ranking criteria for all disease-treatment pairs were generated offline by the \texttt{Qwen2.5-7B-Instruct} model.

\subsubsection{Evaluation Metrics}
\noindent
In our experiments, we employed the Normalized Discounted Cumulative Gain (NDCG)@10 (N@10) and Recall@10 (R@10) as our principal evaluation metrics. 
NDCG@k is a standard listwise accuracy metric that considers the fine-grained relevance and position of documents within the top k positions in the ranked list.
Recall@k, on the other hand, measures the proportion of relevant items ranked within the top-k results, reflecting the model’s ability to cover relevant candidates.

\subsection{Zero-shot Ranking performance}
\subsubsection{Results on DrRank Dataset}
\noindent
To demonstrate the effectiveness of our LLM-based framework, we compare Qwen2.5 series with various baselines, and the results are shown in Table~\ref{table:main}. 
From the results, we obtain several observations.

First, Qwen2.5 models, with scale exceeding 7 billion parameters, consistently outperform all baseline models in terms of NDCG@10 and Recall@10 metrics. This exceptional performance highlights the strong domain generalization capabilities of LLMs, demonstrating the effectiveness of our framework in accurately assessing a doctor's professional relevance in relation to patients' specific medical needs.
As the model size increases, there is a noticeable improvement in the ability to rank doctors accurately, with the 32B model achieving the highest performance, reaching an NDCG@10 of 77.41. Smaller-scale LLMs, however, exhibit relatively poorer performance, underscoring the importance of sufficient model size when addressing complex tasks like doctor ranking in a zero-shot setting.

Second, the introduction of ranking criteria significantly enhances the ranking performance of mid-to-large-scale Qwen2.5 models. Notably, the 3B, 7B, and 32B models all exhibited consistent improvements when ranking criteria were provided, highlighting the strong potential of explicit criteria in guiding LLMs toward more accurate ranking decisions.
However, we observed a contrasting trend in the small model (0.5B and 1.5B): incorporating ranking criteria led to a substantial performance drop, with NDCG@10 decreasing from 55.18 to 45.01 for 0.5B model. 
This indicates that while ranking criteria can enrich the ranking task and potentially boost the upper bound of model performance, they also raise the minimum capability requirements for the model to effectively leverage the additional information.
To further explore the mechanism behind the effectiveness of ranking criteria, we investigated whether the observed improvements stem from the overall format or from the specific disease-treatment-related content. This question is analogous to recent studies on in-context learning, which examine what aspects of demonstrations contribute most to model performance~\upcite{icl_work}. For each disease-treatment pair, we randomly inserted irrelevant but format-consistent ranking criteria into the prompt. We found that such random criteria negatively impacted the performance of the 7B and 32B models, suggesting that it is not merely the presence of structured information, but the actual semantic relevance of the ranking criteria, that plays a key role in guiding the ranking decisions of LLMs.

\begin{figure*}[!t]
\centerline{\includegraphics[width=\textwidth]{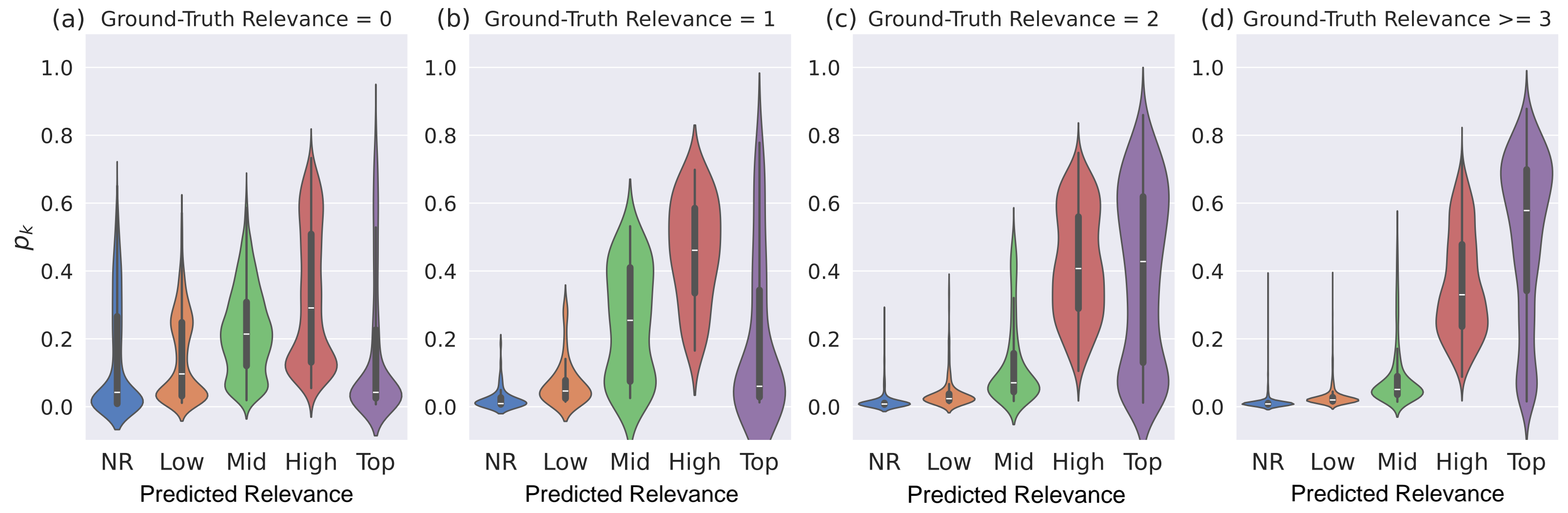}}
\caption{Distribution of marginal probability $p_k$ of each graded relevance label in the aforementioned 5-level setting for query-doctor pairs with different ground-truth labels on the DrRank dataset.
``NR" is an abbreviation for ``Not Relevant".
}
\label{fig:score_distribution}
\end{figure*}

Third, we evaluated the performance of Baichuan-M1-14B-Instruct, a domain-specialized LLM tailored for medical scenarios. By incorporating disease-specific ranking criteria, its NDCG@10 score improved from 71.93 to 74.67, reflecting a notable 4\% gain, in contrast to only a 1\% improvement observed in other models. This result highlights the potential of domain-specific medical knowledge to enhance the effectiveness of ranking criteria, underscoring the promise of explainable ranking in clinical applications of medical LLMs.
However, research on applying medical LLMs to complex clinical ranking tasks remains limited. Future work is needed to better integrate domain expertise with task-specific generalization, paving the way for more robust and interpretable ranking systems in the medical domain.

\subsubsection{Intuitive Score Distribution}
\noindent
We group the query-doctor pairs in the DrRank dataset by different ground-truth relevance labels. 
For every group, we denote $p_k$ as the random variable on our graded label $l_k$ of the marginal probability $\mathrm{prob_{i,k}}$ derived for different query-doctor pairs $(q, {d_i})$.
We then plot the estimated distribution of $p_k$ for each graded label $l_k$ respectively.
In Fig.~\ref{fig:score_distribution}, we plotted the score distribution of our framework implemented with the \texttt{Qwen2.5-7B-Instruct} model. 
We observed that the distributions of $p_k$ of graded label ``Not Relevant'', ``Low'', and ``Mid'' shift down towards 0 as the ground truth relevance increases. 
Meanwhile, the distributions of $p_k$ across relevance label ``Top'' shift up towards 1.
This intuitively reveals how our LLM-based pointwise ranker with fine-grained labels works in practice.

\subsection{Discussions}
\begin{table}
\caption{
The performance of using varying numbers of fine-grained relevance labels on the DrRank dataset. 
Here, ``T", ``H", ``M", ``L", and ``NR" are abbreviations for ``Top", ``High", ``Mid", ``Low", and ``Not Relevant", respectively. 
\Checkmark\ indicates the use of generated ranking criteria; \XSolidBrush\ denotes the absence of ranking criteria.
The best performance is highlighted in bold. 
}
\setlength{\tabcolsep}{6pt}
\centering
\begin{tabular}{rccc}
\toprule
Fine-grained Labels & Criteria & N@10 & R@10\\
\midrule
T, H, M, L, NR  & \Checkmark & \textbf{73.73} & \textbf{29.99}  \\
T, H, M, L, NR  & \XSolidBrush & 72.16 & 29.37  \\
H, M, L, NR   & \Checkmark & 66.27 & 29.52  \\
H, M, L, NR  & \XSolidBrush &  48.78 & 26.43  \\
H, L, NR  & \Checkmark & 34.14 & 22.01  \\
H, L, NR  & \XSolidBrush & 16.00 & 14.38  \\
H, NR  & \Checkmark & 26.54 & 19.29 \\
H, NR  & \XSolidBrush & 18.91 & 16.17  \\
\bottomrule
\end{tabular}
\label{table:label_num}
\end{table}

\subsubsection{Number of Relevance Labels}
\noindent
We investigate the effect of utilizing varying numbers of fine-grained labels on the performance of the \texttt{Qwen2.5-7B-Instruct} model. 
As shown in Table \ref{table:label_num}, increasing the number of relevance labels results in improved performance. 
This enhancement could be attributed to the fine granularity division in the DrRank dataset, which necessitates providing the model with a maximum number of fine-grained labels to facilitate differentiation. 
Additionally, we observed that the ranking criteria consistently demonstrate positive effects across different numbers of fine-grained labels.

\begin{table}
\caption{The performance of each model with ranking criteria using different strategies for deriving ranking scores on the DrRank dataset.
In each experimental group of strategies, the best performance is highlighted in bold.
}
\setlength{\tabcolsep}{6pt}
\centering
\begin{tabular}{lccc}
\toprule
Model & Strategy & N@10 & R@10 \\
\midrule
Qwen2.5-7B-Instruct & $\mathrm{SUM}$ & 73.73 & 29.99 \\
Qwen2.5-7B-Instruct & $\mathrm{MAX_{logit}}$ & 72.15 & \textbf{30.03} \\
Qwen2.5-7B-Instruct & $\mathrm{MAX_{prob}}$ & \textbf{73.78} & 30.02 \\

Qwen2.5-32B-Instruct & $\mathrm{SUM}$  & 77.41 & 30.73 \\
Qwen2.5-32B-Instruct & $\mathrm{MAX_{logit}}$  & 75.97 & \textbf{30.93} \\
Qwen2.5-32B-Instruct & $\mathrm{MAX_{prob}}$ & \textbf{77.44} & 30.73 \\
\bottomrule
\end{tabular}
\label{table:score_strategy}
\end{table}

\subsubsection{Ranking Score Derivation}
\noindent
In our proposed framework, we assign scores to graded labels and calculate the expected value to derive the final score, which we name $\mathrm{SUM}$.
As described in~\upcite{pointwise2score}, another straightforward strategy involves using solely the logit of the highest relevance label, i.e.  ``Top",  which we denote as $\mathrm{MAX_{logit}}$.
Alternatively, we can employ the marginal probability of the highest relevance label after applying the softmax function as \eqref{eq2}, which we refer to as $\mathrm{MAX_{prob}}$.
Table \ref{table:score_strategy} presents the performance of various models with ranking criteria when using different strategies for deriving ranking scores.
We found that the models’ performance fluctuates slightly across three scoring strategies, yet remains consistently high.
This suggests that the improvement from scoring fine-grained relevance labels cannot be simply attributed to the expected summation approach, which is essentially equivalent to multiple sampling.
This conclusion aligns with the findings reported in~\upcite{pointwise2score}.
Future work could focus on developing more effective scoring strategies to further enhance the ranking performance of LLMs.

\subsubsection{Fairness Considerations}
\noindent
The doctor ranking system represents a highly sensitive application domain, which can help patients make important and life-changing decisions.
It is crucial to ensure that these decisions do not exhibit discriminatory behavior towards specific groups or populations~\upcite{fairness}.
To assess the fairness of our proposed framework, we evaluate it from three perspectives: diseases, patient gender, and patient regions.
For disease fairness, we selected 10 diseases with adequate numbers of candidate doctors.
Through proportional sampling based on ground truth labels, we obtained 25 doctors for each disease, ensuring 5 doctors per ground truth label.
This approach helps to standardize the difficulty of doctor ranking across diseases as much as possible. 
We computed the standard deviation of the NDCG@10 metric among the various diseases to measure the model's bias. 
This process was repeated 1,000 times, and the average of these standard deviations was taken as the final result.
To evaluate fairness related to patient gender and regions, we prepended sensitive information such as ``I am male/female" or ``I come from an urban/rural area" to the query. 
If the model displays discriminatory tendencies based on gender or region, these biases will be reflected in the NDCG@10 metric.

\begin{table}
\caption{The analysis results for fairness in diseases, patient gender, and patient regions on the DrRank dataset. 
``SD" is an abbreviation for ``standard deviation".
``bge-reranker" refers to ``bge-reranker-v2-m3".
``Qwen-7B" refers to ``Qwen2.5-7B-Instruct", and the case with ranking criteria is marked with *.
}
\setlength{\tabcolsep}{6pt}
\centering
\begin{tabular}{lccccc}
\toprule
Model & Disease & \multicolumn{2}{c}{Gender} & \multicolumn{2}{c}{Region} \\
& SD & \multicolumn{2}{c}{NDCG@10} & \multicolumn{2}{c}{NDCG@10} \\
\cmidrule(lr){3-4} \cmidrule(lr){5-6}
& & male & female & urban & rural \\
\midrule
bge-reranker & 0.0773 & 66.73 & 60.21 & 60.21 & 59.96\\
Qwen-7B & 0.0689 & 72.68 & 73.96 & 73.96 & 74.31\\
Qwen-7B* & 0.0672 & 73.26 & 73.98 & 73.98 & 75.20\\
\bottomrule
\end{tabular}
\label{table:fairness}
\end{table}

Table \ref{table:fairness} presents the analysis results for fairness concerning diseases, patient gender, and patient regions on the DrRank dataset.
We observed that our LLM-based frameworks exhibit superior performance in terms of disease fairness. 
Their standard deviations (SD) were 0.0689 and 0.0672, respectively, which are lower than the 0.0773 SD of bge-reranker-v2-m3. 
This indicates that our models' performance is more consistent across different diseases. 
We observe that the introduction of ranking criteria further enhances fairness across disease scenarios by providing standardized guidelines, as reflected by the SD decreasing from 0.0689 to 0.0672.
Regarding gender fairness, our framework demonstrates a significant advantage, yielding very similar NDCG@10 scores for male and female patients, with differences of less than 1.3 and 0.8, respectively.
In contrast, bge-reranker-v2-m3 displays a pronounced gender bias, with its NDCG@10 score difference between genders reaching 6.52.
As for regional fairness, all evaluated models performed relatively similarly for urban and rural patients, indicating comparable levels of fairness in this dimension.
Although our frameworks show improvements in fairness, further research remains crucial for ensuring and enhancing the fairness of LLMs in ranking tasks, particularly in addressing biases like the one observed in the baseline model.

\begin{table*}
\caption{Annotation results for the interpretability of ranking criteria and evaluation rationales, rated by five medical experts.}
\label{table:interpretability}
\centering
\begin{tabular}{llcccccc}
\toprule
 & Metric & {Expert 1} & {Expert 2} & {Expert 3}  & {Expert 4}  & {Expert 5} & {Average}\\
\midrule
\multirow{2}{*}{Ranking Criteria} 
& Fluency & 4.85 & 3.95 & 4.97 & 4.23 & 4.36 & 4.47 \\
& Professionalism & 4.56 & 4.69 & 3.87 & 4.52 & 4.14 & 4.36 \\
\cmidrule(l){2-8} 
\multirow{2}{*}{Evaluation Rationales} 
& Fluency & 4.63 & 4.65 & 4.44 & 4.55 & 4.58 & 4.57 \\
& Rationality & 3.87 & 4.39 & 4.23 & 4.12 & 3.98 & 4.12 \\
\bottomrule
\end{tabular}
\end{table*}

\begin{table*}
\caption{Quantization Impact on \texttt{Qwen2.5-32B-Instruct} Performance. 
\Checkmark\ indicates the use of generated ranking criteria; \XSolidBrush\ denotes the absence of ranking criteria.
``Scoring Time'' and ``Explaining Time'' denote the average inference time to produce relevance scores and natural-language explanations, respectively. GPU memory is the per-GPU peak usage across four A6000 cards. 
}
\centering
\setlength{\tabcolsep}{4pt}
\begin{tabular}{lcccccc}
\toprule
Model & Criteria & Scoring Time (s) & Explaining Time (s) & GPU Memory (GB) & N@10 & R@10 \\
\midrule
Qwen2.5-32B-Instruct            & \Checkmark        & 0.10                       & 60.23                        & 29.85                   & 77.41            & 30.73           \\
Qwen2.5-32B-Instruct            & \XSolidBrush      & 0.10                       & 36.42                        & 29.85                   & 76.61            & 30.40           \\
Qwen2.5-32B-Instruct-AWQ        & \Checkmark        & 1.19                       & 42.15                        & 19.35                   & 76.95            & 30.72           \\
Qwen2.5-32B-Instruct-AWQ        & \XSolidBrush      & 0.84                       & 24.92                        & 19.35                   & 75.59            & 30.71           \\
\bottomrule
\end{tabular}
\label{table:quantization}
\end{table*}

\subsubsection{Interpretability Considerations}
\noindent
Our framework improves the interpretability of doctor ranking by incorporating ranking criteria as standard evaluation guidelines and by generating evaluation rationales to serve as recommendation explanations.
Although LLMs excel in text generation and logical reasoning, they occasionally produce hallucinations~\upcite{hallucination}, which undermine the rationality and interpretability of medical advice.
To assess our ranking criteria and evaluation rationales, we specifically engaged five medical experts.
The ranking criteria were evaluated based on fluency and professionalism.
``Fluency" refers to the LLM-generated text being smooth, coherent, and without grammatical or logical errors.
``Professionalism" indicates that the LLM-generated ranking criteria comprehensively reflect doctors' professional perspectives on a specific disease-treatment pair.
The evaluation rationales were assessed based on fluency (same as above) and rationality.
``Rationality" signifies that the LLM-generated evaluation rationales are medically logical and reasonable.
All assessments were scored on a 0 to 5 point scale, with higher scores indicating higher recognition.
We used the \texttt{Qwen2.5-7B-Instruct} model to generate these criteria and rationales.
In total, 38 ranking criteria for various disease-treatment pairs, and 140 evaluation rationales for 28 disease-treatment pairs (with 5 doctors each) were annotated.
The high scores detailed in Table \ref{table:interpretability} validate the output of our framework, with medical experts rating both ranking criteria and evaluation rationales favorably for fluency, professionalism, and rationality. 
This provides strong evidence of the framework's potential to address the interpretability challenge in doctor ranking. 
To complement this expert-based assessment, our immediate next step is to engage patients in user studies to gather authentic experiential feedback.

\subsubsection{Low-Bit Quantization for Deployment}
\noindent
Quantization refers to reducing model weight precision (i.e., the number of bits) with minimal performance loss during inference~\upcite{Quantization}. 
This procedure significantly lowers GPU memory requirements, and when combined with low-precision hardware and operator optimizations, can further accelerate inference.
As a representative weight-only low-bit quantization approach, AWQ~\upcite{AWQ} retains the top 1\% of weights—those most critical to LLM performance—in high precision, and applies a per-channel scaling mechanism to determine optimal scale factors.
To evaluate our framework under resource-constrained conditions, we employ \texttt{Qwen2.5-32B-Instruct-AWQ}\footnote{\url{https://huggingface.co/Qwen/Qwen2.5-32B-Instruct-AWQ}} via the AutoAWQ library\footnote{\url{https://github.com/casper-hansen/AutoAWQ}} and measure the impact of quantization on the DrRank dataset. Table~\ref{table:quantization} reports the average inference times (for scoring and explaining), per-GPU memory footprint (four A6000 GPUs), and ranking quality.
The results demonstrate that AWQ quantization achieves a 35.2\% reduction in GPU memory (from 29.85 GB to 19.35 GB) while maintaining comparable ranking performance (NDCG@10 drops from 77.41 to 76.95; Recall@10 remains effectively unchanged). Explaining time is reduced by approximately 30\%, whereas scoring time increases by nearly 10×. 
We attribute this substantial scoring-time overhead to the cost of bit-packing and unpacking operations during the LLM’s prefill stage, which, absent optimized kernels, impose significant latency. 
Overall, quantization offers promise for deploying our framework in resource-constrained environments.

\begin{table}[h!]
\caption{Generalization performance (NDCG@10) on two datasets from BEIR benchmark. Model with ranking criteria is marked with *. The best performance is highlighted in bold.}
\centering
\setlength{\tabcolsep}{8pt} 
\renewcommand{\arraystretch}{1.1} 
\begin{tabular}{llc}
\toprule
Dataset & Model & NDCG@10 \\
\midrule
\multirow{3}{*}{\texttt{TREC-COVID}} & bge-reranker-v2-m3 & 81.79 \\
& Qwen2.5-7B-Instruct & 85.27 \\
& Qwen2.5-7B-Instruct* & \textbf{89.46} \\
\midrule
\multirow{3}{*}{\texttt{Touche-2020}} & bge-reranker & 76.74 \\
& Qwen2.5-7B-Instruct & 82.66 \\
& Qwen2.5-7B-Instruct* & \textbf{88.88} \\
\bottomrule
\end{tabular}
\label{table:generalization}
\end{table}

\subsubsection{Robustness and Generalization}
\noindent
To validate that our framework is not a niche solution tailored only for medical expert finding, we evaluate its performance on broader information retrieval tasks. We conduct a zero-shot generalization experiment on two datasets from the BEIR benchmark: \texttt{trec-covid} (scientific literature retrieval) and \texttt{touche2020} (argument retrieval).
We compare our LLM-based ranker (based on \texttt{Qwen2.5-7B-Instruct}) against the \texttt{bge-reranker-v2-m3} baseline. The results, presented in Table~\ref{table:generalization}, demonstrate the great robustness of our approach. 
On both datasets, our LLM-based ranker, even without ranking criteria, significantly outperforms the baseline method. This highlights the superior contextual reasoning capability of LLMs for complex ranking tasks compared to interaction-based methods.
Furthermore, a second, consistent performance increase is achieved when our dynamically generated ranking criteria are incorporated. This again confirms that providing the LLM with a structured, task-adaptive ranking criteria is a key factor in boosting its ranking accuracy and reliability.
Overall, by making both the decision-making process (via the criteria) and the final outputs (via generated rationales) transparent, our framework offers a high-performing and trustworthy solution for a wide range of ranking tasks.


\subsection{Limitations}
\noindent
While our proposed framework exhibits strong zero-shot generalization and interpretability in ranking doctors based on professional relevance, it remains primarily a research-focused solution.
Practical constraints—such as doctor availability, appointment quotas, patient insurance coverage, and the real-time status of healthcare platforms—are not currently addressed.
Future work could explore integration with online medical service providers to incorporate real-time scheduling and patients’ financial considerations, thereby improving real-world applicability.
Another limitation lies in the potential for exacerbating patient distribution imbalances: experienced doctors with strong credentials may consistently receive high relevance scores, leading to recommendation overload.
However, mitigating such exposure bias is beyond the scope of this study, which focuses solely on professional relevance.
Future research could investigate diversity-aware and workload-aware objectives to improve system sustainability.

\begin{figure*}
\centerline{\includegraphics[width=\textwidth]{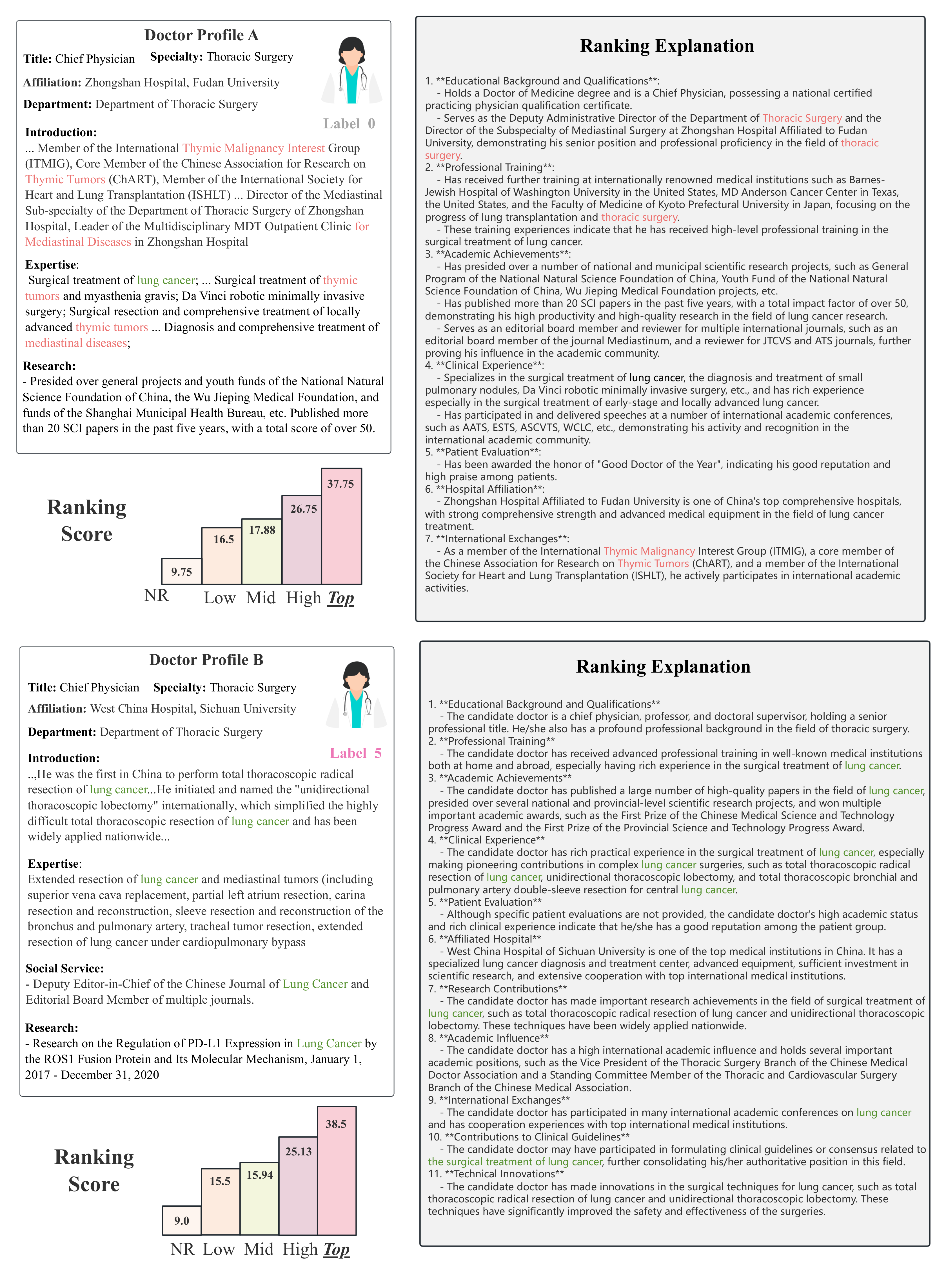}}
\caption{A comparative case study illustrating the ranking outcomes of our LLM-based doctor ranking framework, where the patient query is: ``I want to find a doctor specializing in surgical treatment for lung cancer.''
}
\label{fig:case}
\end{figure*}

\section{Case Study}
\noindent
To further illustrate the capabilities and limitations of our proposed framework, we present a comparative case study involving two thoracic surgeons—Doctor A and Doctor B, as shown in Fig.~\ref{fig:case}. This case highlights how the model interprets nuanced professional profiles.
The patient query in this case is: ``I want to find a doctor specializing in surgical treatment for lung cancer." The two candidate doctors possess rich professional backgrounds and are affiliated with top-tier hospitals in China. Expert annotations label Doctor B with the highest relevance score of 5, while Doctor A receives a relevance score of 0 from hard negative mining (Section \ref{sec:hnm}).
Our LLM-based ranker predicts high professional relevance for both candidates, assigning similarly high scores. The model’s top-1 predicted label for both doctors is ``Top", and the accompanying explanations are detailed and well-structured, covering dimensions such as clinical expertise, academic influence, international exchange, and institutional affiliation. Both outputs demonstrate the framework’s ability to extract and reason over structured and unstructured data.

Despite the coherent reasoning, a critical misjudgment occurred in the case of Doctor A. While the doctor's profile is comprehensive and impressive, their core expertise centers around \textit{mediastinal tumors} and \textit{thymic diseases}, rather than the \textit{lung cancer surger}. The profile mentions lung cancer treatment only as a secondary area of focus, and the doctor is not known as a top expert in this domain. However, the model overemphasized: (1) The institutional prestige (Zhongshan Hospital); (2) Rich international academic exchange; (3) General thoracic surgical achievements. This suggests the model may sometimes conflate general thoracic surgery excellence with disease-specific relevance, especially when abundant textual signals (e.g., awards, fellowships, academic positions) overwhelm more subtle cues related to disease-treatment alignment. In contrast, Doctor B is correctly identified as a top expert. The model emphasized key attributes such as: (1) Pioneering minimally invasive surgical procedures for lung cancer; (2) Contributions to national clinical guidelines related to the surgical treatment of lung cancer; (3) High-volume, high-difficulty surgical experience in central-type lung cancer; (4) Leading roles in national medical associations and top-tier journals in the field of surgical treatment of
lung cancer. These signals matched expert annotations and demonstrated the model's strength in handling multi-dimensional relevance criteria when such signals are clearly aligned with the disease-treatment query.

This case demonstrates that how \textit{false positives} in recommender systems can lead to suboptimal doctor-patient parings. 
Although Doctor A is highly accomplished, patients seeking specialized care for lung cancer surgery might receive recommendations that are not optimally aligned with their medical needs. 
This reflects a broader challenge of semantic drift in zero-shot settings, where strong but tangential signals bias the ranking outcome. 
It also highlights the importance of grounding disease-treatment relevance beyond general academic or clinical excellence. 
Acknowledging this issue, we identify a valuable path for future improvement: using doctors' publication topics as a form of automated post-hoc calibration. The specific focus of a doctor's published research provides a powerful signal of their nuanced specialty. Although not incorporated in our current evaluation due to resource limitations, this data source is crucial for correcting model bias and grounding recommendations in highly specific expertise.

\section{Conclusion}
\noindent
This study presents a zero-shot, explainable doctor ranking framework powered by large language models, aiming to address critical challenges in data scarcity, interpretability, and expertise alignment in medical expert finding. 
The proposed approach integrates fine-grained pointwise relevance estimation, dynamic generation of ranking criteria, and natural language-based explanations, achieving strong performance while ensuring notable transparency and fairness on the new, expertise-driven DrRank dataset.
Furthermore, we demonstrate the framework’s robust generalization on two datasets from BEIR benchmark, confirming its effectiveness as a general-purpose, explainable ranking framework.
Future work will focus on further leveraging domain-specific medical knowledge—such as academic publications and clinical records—to enrich the ranking process, as well as exploring the integration of our LLM-based framework with real-world commercial healthcare systems to ensure practical applicability.


\vskip 2mm
\zihao{5}
\noindent
\textbf{Acknowledgment}
\vskip 2mm

\zihao{5--}
\noindent
This study was funded by the National Natural Science Foundation of China (62203060, 62403492), R\&D Program of Beijing Municipal Education Commission (KM202310005030), the National High Level Hospital Clinical Research Funding (2022-PUMCH-C-017).

\vskip 2mm
\renewcommand\refname{\zihao{5}\textbf{References}}

\begingroup
\zihao{5-}
\bibliographystyle{ieeemod}
\bibliography{references}

\endgroup

\begin{strip}
\end{strip}

\begin{biography}[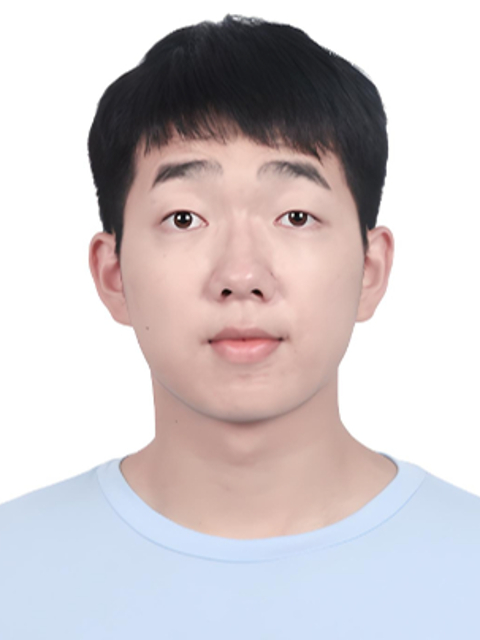]
\noindent
\textbf{Ziyang Zeng} received the BEng degree in Computer Science and Technology from Beijing University of Posts and Telecommunications, China, in 2023. 
He is currently a Master's student at the School of Information and Communication Engineering, Beijing University of Posts and Telecommunications. 
His research interests include large language models, information retrieval, and reinforcement learning.
\end{biography}

\begin{biography}[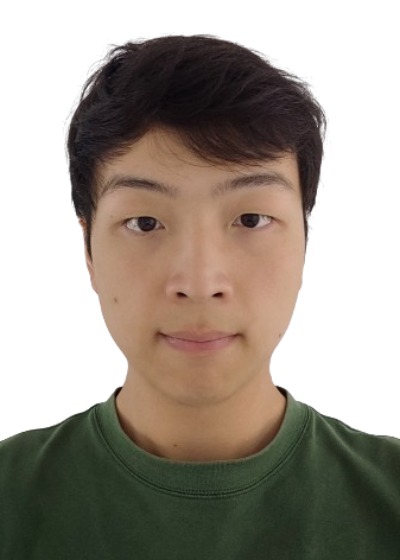]
\noindent
\textbf{Dongyuan Li} received the BEng degree in Artificial Intelligence at the School of Artificial Intelligence, Beijing University of Posts and Telecommunications, China in 2025. 
He is currently a Master's student at the School of Artificial Intelligence, Beijing University of Posts and Telecommunications. 
His research interests include large language models and multi-agent systems.
\end{biography}

\begin{biography}[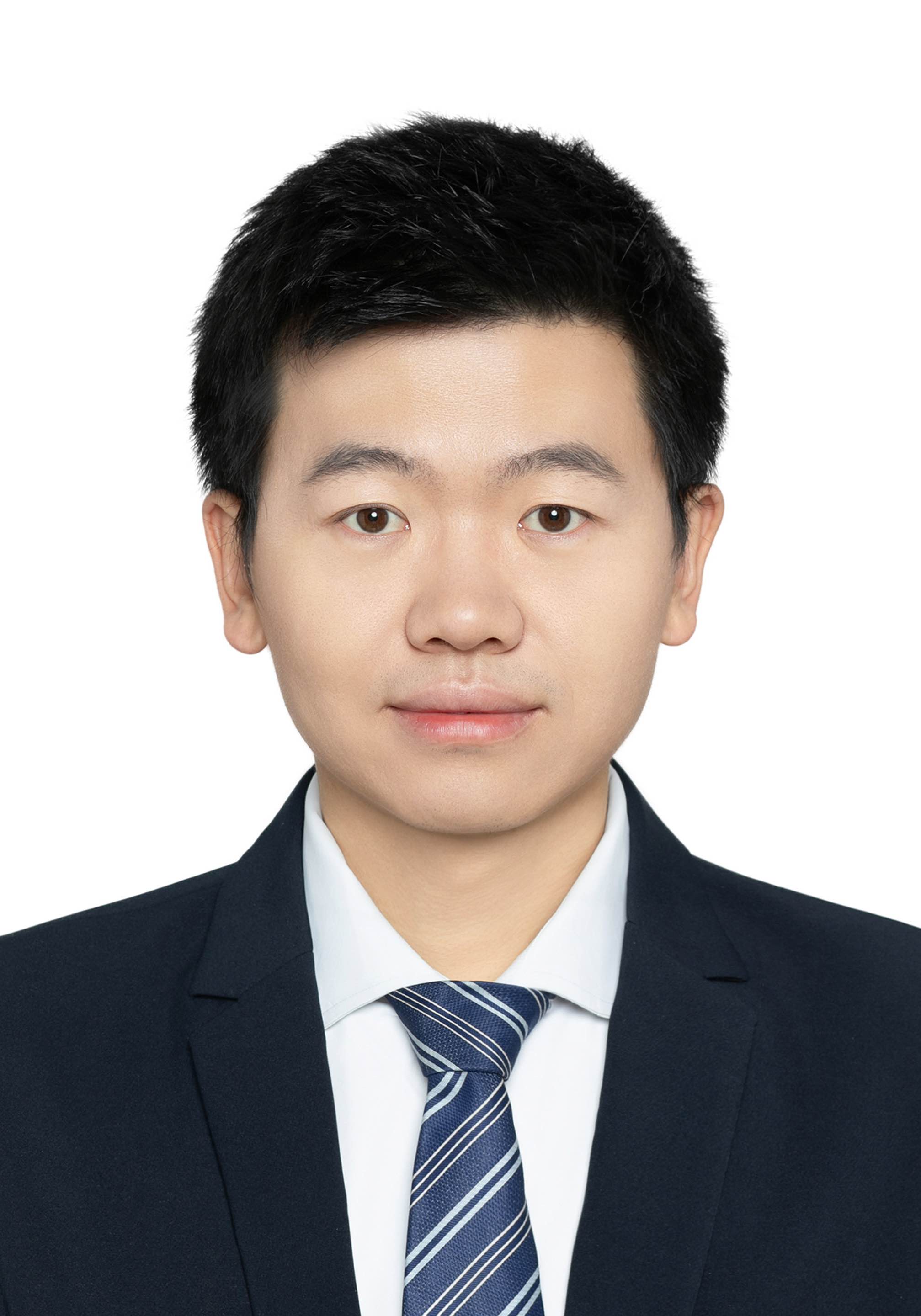]
\noindent
\textbf{Yuqing Yang} received the PhD degree in computer science and technology from Tsinghua University, China in 2019. He joined the Beijing University of Posts and Telecommunications at 2021 and his research interests include artificial intelligence, smart healthcare, and bioinformatics.
\end{biography}

\end{document}